\documentclass{article}
\usepackage[preprint]{neurips_2026}

\usepackage[utf8]{inputenc} 
\usepackage[T1]{fontenc}    
\usepackage{hyperref}       
\usepackage{url}            
\usepackage{booktabs}       
\usepackage{graphicx}
\usepackage{subcaption}
\usepackage{amsmath,amssymb,amsfonts,amsthm}
\usepackage{mathtools}
\usepackage{bbm}       
\usepackage{nicefrac}       
\usepackage{microtype}      
\usepackage{xcolor}         
\usepackage{algorithm}
\usepackage{algorithmic}
\usepackage{cleveref}

\hypersetup{
    colorlinks=true,
    linkcolor=blue,
    citecolor=blue,
    urlcolor=blue
}

\newtheorem{theorem}{Theorem}

\newtheorem{assumption}{Assumption}

\newcommand{\E}{\mathbb{E}}
\newcommand{\Pbb}{\mathbb{P}}
\newcommand{\R}{\mathbb{R}}
\newcommand{\1}{\mathbbm{1}}
\newcommand{\X}{\mathcal{X}}
\newcommand{\T}{\mathcal{T}}
\newcommand{\M}{\mathcal{M}}
\newcommand{\Z}{\mathcal{Z}}
\newcommand{\PiClass}{\Pi}
\newcommand{\argmax}{\operatorname*{argmax}}
\newcommand{\argmin}{\operatorname*{argmin}}

\newcommand{\safe}{\mathrm{safe}}
\newcommand{\fair}{\mathrm{fair}}

\newcommand{\maxth}{\mathrm{max}}

\title{Strategic Causal Policy Learning with Manipulable Eligibility}

\author{%
  Suhwan~Bong\\
  Harvard University\\
  \texttt{suhwanbong@g.harvard.edu} \\
}

\begin{document}

\maketitle

\begin{abstract}
The standard rule in unconstrained causal policy learning is simple: treat individuals with nonnegative CATEs. But this rule can fail when treatment eligibility is strategically manipulable. After a policy is announced, individuals may change the covariates used for assignment without changing their underlying treatment effects. We study how such responses affect treatment allocation and welfare. Using an access-burden representation, we characterize strategic treatment and decompose welfare loss into missed beneficial treatment and harmful treatment. For linear CATEs and quadratic manipulation costs, we derive closed-form threshold corrections. When manipulation costs are homogeneous, a single correction recovers the ideal positive-CATE allocation. With heterogeneous costs, welfare maximization, pointwise safety, and minimax group welfare regret can favor different thresholds. Simulations illustrate these tradeoffs and examine the effects of estimating treatment effects and behavioral responses.
\end{abstract}

\section{Introduction}
\label{sec:introduction}
A common starting point in causal policy learning is to treat individuals whose conditional average treatment effects (CATEs) are nonnegative. In the absence of constraints, this rule maximizes the expected benefit of treatment when the covariates used for assignment are fixed. A large literature studies how to learn such treatment rules from experimental or observational data, including statistical treatment rules, empirical welfare maximization, and policy learning with observational data \citep{manski2004statistical,kitagawa2018should,athey2021policy}. In these settings, policies are evaluated by the treatment gains they assign at observed baseline covariates.

In many policy settings, however, treatment eligibility is strategically manipulable. After an eligibility rule is announced, individuals may change, improve, or report covariates in order to receive treatment. This type of behavior is central in strategic classification, where individuals manipulate features to obtain favorable decisions \citep{hardt2016strategic, dong2018strategic, chen2020learning}. Recent work further emphasizes that strategic responses may be either pure gaming, which changes the decision without improving the underlying outcome, or productive investment, which changes outcome-relevant states \citep{kleinberg2020classifiers, miller2020strategic, shavit2020causal}. 

This distinction matters for causal policy learning.
Under pure gaming, the institution assigns treatment using covariates observed after strategic response, but the welfare value of treatment is determined by the individual's baseline treatment effect. Thus, a person may enter the treatment region without becoming a better treatment candidate, creating a mismatch between who receives treatment and who benefits from treatment. This differs from conventional strategic classification, which typically seeks accurate predictions of a target label under manipulation, and from standard causal policy learning, where treatment assignment is based on fixed baseline covariates.

We formalize this problem as \emph{strategic causal policy learning}. We consider pure gaming, where individuals can change the covariates used for treatment assignment without changing their potential outcomes. We study how manipulation costs affect who receives treatment and the resulting policy value. It also distinguishes our setting from performative prediction, where deployed predictors induce distribution shift \citep{perdomo2020performative}, and from strategic policy learning models in which agents' responses may alter outcomes or equilibrium distributions \citep{sahoo2025policy}. In our setting, strategic response changes access to treatment, while welfare is evaluated using baseline CATEs.

\paragraph{Motivating example.}
Consider a lender or public agency that offers a subsidized credit-building program based on applicants' predicted treatment benefits. Applicants may take costly steps to change the credit information used to determine eligibility. For example, becoming an authorized user on an established credit-card account may improve an applicant's credit history without necessarily changing their underlying financial circumstances. Such actions may involve fees, time, or access to someone willing to share an account. If these changes affect the eligibility rule but leave the program's treatment effect unchanged, an applicant may become eligible without becoming a better candidate for the program.

In the credit-building example, an applicant who can improve their observed credit record at low cost may become eligible even when the program's treatment effect is small or negative. Making eligibility more restrictive could prevent such applicants from receiving treatment, but it may also exclude applicants who would benefit from the program, especially those who cannot easily change their records. When manipulation costs differ across groups, the choice of eligibility rule may also affect how welfare losses are distributed. These tradeoffs raise different policy objectives: maximizing aggregate treatment gains, avoiding harmful treatment, and limiting group-level welfare losses. As we show below, these objectives can lead to different thresholds in a simple linear–quadratic model.

\subsection{Our contributions}
\label{subsec:contributions}

We make four contributions. First, we formalize a model of \emph{strategic causal policy learning}, where treatment is assigned using strategically manipulated covariates but its value is determined by baseline treatment effects. Second, we introduce an access-burden representation and derive a manipulation-gap identity showing how strategic response changes policy value. We also decompose welfare regret into missed beneficial treatment and harmful treatment. Third, for linear CATE models with quadratic manipulation costs, we derive closed-form characterizations of strategic treatment assignment. Homogeneous costs admit a single threshold correction that recovers the ideal positive-CATE allocation, whereas heterogeneous costs can lead to different policy choices depending on the objective, including welfare, safety, and group welfare regret. Fourth, we study what behavioral response data reveal about manipulation buffers and how errors in CATE and response estimation affect policy performance. We derive conditional safety and regret bounds and examine these errors through numerical stress tests.

\subsection{Related work}
\label{subsec:related_work}
\paragraph{Strategic adaptation.}
Strategic adaptation studies decision rules deployed in environments where individuals can manipulate, report, or change features in response to the rule in order to obtain favorable decisions. A central example is strategic classification, where individuals modify features to receive a favorable prediction or decision \citep{hardt2016strategic, dong2018strategic, chen2020learning}. Related work emphasizes that such feature changes may represent either pure gaming or productive improvement, making strategic adaptation inherently causal \citep{kleinberg2020classifiers, miller2020strategic, shavit2020causal}. Strategic responses also arise in treatment allocation, competing-agent policy learning, and performative prediction, where deployed rules induce behavioral responses or distributional change \citep{munro2025treatment, sahoo2025policy, perdomo2020performative, chen2024practical}. \citep{munro2025treatment} studies treatment allocation with strategic agents, allowing randomized policies and developing a sequential experimental approach. We focus on deterministic eligibility rules and use cost-based reachability to study their implications for welfare, safety, and group welfare regret.

\paragraph{Causal policy learning.}
Causal policy learning studies how to assign treatments using experimental or observational data. Various methods have been proposed to estimate optimal treatment \citep{zhang2012robust, zhang2013robust, kallus2018balanced}. Classical work on statistical treatment rules and empirical welfare maximization studies how sample information can guide treatment choice in heterogeneous populations \citep{manski2004statistical, kitagawa2018should}. More recent work develops policy learning methods for observational data and constrained policy classes \citep{athey2021policy}. Fairness-aware policy learning further studies how welfare maximization interacts with group-level constraints and heterogeneous treatment effects \citep{kim2023fair}. Our work builds on this literature and studies how policy learning changes when the covariates used for assignment are strategically manipulated at deployment.

\paragraph{Fairness in strategic settings.}
Fairness in strategic settings often studies group-dependent costs, social burden, and post-response disparities. 
Prior work shows that heterogeneous manipulation costs can produce disparate effects across groups and that strategy-robust classification can impose social burdens \citep{hu2018short,milli2019social, diana2025minimax, keswani2023addressing, estornell2021unfairness, braverman2020role}. 
The recourse literature similarly studies whether individuals can change actionable features to obtain a desired decision, often through minimum-cost changes to a decision boundary \citep{wachter2017counterfactual,ustun2019actionable}.

\section{Model}
\label{sec:model}
\subsection{Strategic Causal Policy Learning}
Let $X\in\X$ denote baseline covariates. Treatment is binary, $A\in\{0,1\}$, with potential outcomes $Y(1)$ and $Y(0)$. The conditional average treatment effect (CATE) is
\[
  \tau(x)=\E[Y(1)-Y(0)\mid X=x].
\]
We assume that the CATE is identified from available data under standard causal identification conditions, such as consistency, overlap, and unconfoundedness conditional on $X$. We view CATE estimation as a first-stage problem and focus on policy learning under strategic response.

For any assignment rule $a:\X\to\{0,1\}$, the treatment-gain component of welfare is $\E[\tau(X)a(X)]$. Full outcome welfare also contains the policy-independent term $\E[Y(0)]$, which we omit.

A deterministic policy is a map $\pi:\X\to\{0,1\}$. Let $\PiClass$ denote the class of policies available to the institution. After the institution announces $\pi$, an individual with baseline covariates $X=x$ can report or move to $x'$. Treatment is assigned using the post-response covariates, $\pi(x')$, rather than $\pi(x)$.

\begin{assumption}\label{ass:pure_gaming}
Strategic changes to the covariates used for assignment do not directly change potential outcomes. If $X^\pi$ denotes the post-response covariates under $\pi$, then the realized outcome is $Y(\pi(X^\pi))$.
\end{assumption}

The strategic treatment-gain objective is therefore
\begin{equation}
  G^S(\pi)=\E[\tau(X)\pi(X^\pi)].
  \label{eq:strategic_gain_general}
\end{equation}

\subsection{Strategic Response and Treatment Assignment}
Let $\mathcal A(x)\subseteq\X$ be the set of feasible reports from baseline $x$, with $x\in\mathcal A(x)$. Let $c(x',x)\ge0$ be the cost of choosing $x'$, with $c(x,x)=0$. Individuals choose a best response according to
\begin{equation}
  X^\pi\in\argmax_{x'\in\mathcal A(X)}\{\pi(x')-c(x',X)\}.
  \label{eq:best_response}
\end{equation}
The unrestricted-action case corresponds to $\mathcal A(x)=\X$; more generally, $\mathcal A(x)$ can restrict which features may be changed.

The unit benefit of receiving treatment in this best-response problem is a private incentive, not the treatment gain measured by the CATE. Thus a person with a negative CATE may still prefer treatment. Manipulation costs govern behavior but are not subtracted from the institutional treatment-gain objective. We assume that a best response exists and that ties between treated and untreated reports are resolved in favor of treatment.

\subsection{Access Burden and Strategic Treatment}
Let $\T_\pi=\{x'\in\X:\pi(x')=1\}$ be the treatment region. The \emph{access burden} is the minimum cost of reaching that region from baseline $x$:
\begin{equation}
  b_\pi(x)=\inf_{x'\in\mathcal A(x)\cap\T_\pi}c(x',x),
  \label{eq:access_burden}
\end{equation}
where $b_\pi(x)=\infty$ if no treated report is feasible. We assume that this infimum is attained whenever $b_\pi(x)=1$.

The best-response rule and treatment-favoring tie convention imply
\begin{equation}
  \pi(X^\pi)=\1\{b_\pi(X)\le1\}.
  \label{eq:access_representation}
\end{equation}

Define the strategic treatment indicator $m_\pi(x)=\1\{b_\pi(x)\le1\}$. The strategic value can now be written as
\begin{equation}
  G^S(\pi)=\E[\tau(X)m_\pi(X)].
  \label{eq:strategic_value_access}
\end{equation}
An individual contributes their baseline treatment effect to this value if and only if they can reach the treatment region at cost at most one.

\paragraph{Running example.} In the credit-building example, baseline characteristics $X$ include information used to assess an applicant's treatment benefit, while $X^\pi$ contains the eligibility information after the applicant responds. The access burden $b_\pi(x)$ is the least cost required to produce an eligible credit-file report. Differences in access can be reflected in the feasible actions and their costs.

\section{Policy Evaluation under Strategic Response}
\label{sec:policy_evaluation}
\subsection{Strategic Welfare and Regret}
In the absence of strategic response, the policy is applied directly to baseline covariates. The non-strategic value of a policy is therefore
\begin{equation*}
  G^N(\pi)=\E[\tau(X)\pi(X)].
\end{equation*}
The next identity shows exactly how strategic response changes this value.

\begin{theorem}[Manipulation-gap identity]
\label{thm:manipulation_gap}
For any policy \(\pi\),
\begin{equation}
  G^S(\pi)-G^N(\pi)
  =
  \E\!\left[\tau(X)\1\{\pi(X)=0,\ b_\pi(X)\le1\}\right].
  \label{eq:manipulation_gap}
\end{equation}
\end{theorem}
Let $\M_\pi=\{x:\pi(x)=0,\ b_\pi(x)\le1\}$ be the manipulation band. Then,
\begin{equation*}
  G^S(\pi)-G^N(\pi)=\E[\tau(X)\1\{X\in\M_\pi\}].
\end{equation*}
The gap is positive when the manipulation band contains more positive-CATE mass than negative-CATE mass, and negative when harmful entrants dominate.

The ideal policy treats individuals with positive treatment effects,
achieving $G^\star=\E[\tau(X)_+]$.
A strategic policy can fall short of this benchmark in two ways:
it may leave individuals with positive treatment effects untreated
or assign treatment to those with negative treatment effects.
Accordingly,
\begin{equation*}
  G^\star-G^S(\pi)
  =
  \E[\tau(X)_+\1\{m_\pi(X)=0\}]
  +
  \E[(-\tau(X))_+\1\{m_\pi(X)=1\}].
\end{equation*}
The two terms represent missed beneficial treatment and harmful treatment, respectively.

\subsection{Policy Objectives}
The representation above applies to any deterministic policy in a class $\PiClass$, not only to CATE thresholds. Different institutional objectives may therefore use the same strategic treatment rule in different ways.

\paragraph{Welfare.}
The welfare-maximizing policy solves
\[
  \max_{\pi\in\PiClass}G^S(\pi).
\]
Maximizing aggregate treatment gains need not eliminate
harmful treatment, since individuals with negative CATEs
may receive treatment under strategic response.

\paragraph{Safety.}
An institution may instead wish to limit harmful treatment.
Define
\[
  H(\pi)=\E[(-\tau(X))_+m_\pi(X)].
\]
A safety-constrained policy solves
\[
  \max_{\pi\in\PiClass}G^S(\pi)
  \quad\text{subject to}\quad H(\pi)\le\eta,
\]
where $\eta\ge0$ is the permitted level of harm.
Setting $\eta=0$ rules out harmful treatment almost surely.
A stronger pointwise safety requirement excludes treatment
for every $x$ with $\tau(x)<0$.

\paragraph{Group welfare regret.}
We now extend the model to an observed, immutable group
attribute $Z\in\Z$. For group comparisons, assume that
$\Z$ is finite and each group has positive probability.
The institution may condition its announced rule on $Z$,
but individuals cannot manipulate group membership.
Write
\[
  \tau_z(x)=\E[Y(1)-Y(0)\mid X=x,Z=z]
\]
and let $c_z$ and $\mathcal A_z(x)$ denote the corresponding
manipulation cost and feasible reports. The group-specific
strategic treatment indicator is
$m_{\pi,z}(x)=\1\{b_{\pi,z}(x)\le1\}$, where
$b_{\pi,z}(x)$ is the minimum cost of reaching
$\{x':\pi(x',z)=1\}$ from $x$.
The general results above apply conditionally on group
membership. In this extension, aggregate strategic welfare
and its ideal benchmark are
\[
  G^S(\pi)=\E[\tau_Z(X)m_{\pi,Z}(X)],
  \qquad G^\star=\E[(\tau_Z(X))_+].
\]
For each group, define the ideal and strategic values
\[
  B_z^\star=\E[(\tau_z(X))_+\mid Z=z],
  \qquad
  B_z^S(\pi)=\E[\tau_z(X)m_{\pi,z}(X)\mid Z=z].
\]
The group welfare regret is
\[
  F_z(\pi)=B_z^\star-B_z^S(\pi).
\]
One possible fairness objective is to minimize
$\max_{z\in\Z}F_z(\pi)$ over the available policies.
Unlike treatment-rate parity, this criterion compares
each group's causal welfare with its own ideal allocation.

\paragraph{Running example.}
In the credit-building example, applicants may become eligible
by changing their credit records, even if they would not qualify based on their original records. This increases welfare when treatment benefits these applicants but reduces it when treatment is harmful. The manipulation gap captures both possibilities, while safety constraints seek to limit harmful treatment. Group welfare regret also accounts for applicants who would benefit from treatment but remain untreated because qualifying is too costly. Thus, differences in manipulation costs can lead to different welfare losses across groups under the same eligibility rule.

\section{Linear CATEs and Quadratic Costs}
\label{sec:linear_l2}
We specialize the group-conditional extension of
Section~\ref{sec:policy_evaluation} to CATE-threshold
policies with linear treatment effects and quadratic
manipulation costs. Here $\X=\R^d$. Throughout this section, let
$\tau_z(x)=\beta^\top x+\delta_z$, where $\beta\ne0$,
and $c_z(x',x)=\lambda_z\|x'-x\|_2^2/2$, where
$\lambda_z>0$. Reports are unrestricted,
$\mathcal A_z(x)=\R^d$, but group membership is
immutable. Pointwise statements apply to all
$x\in\R^d$.

We consider a common-threshold policy
$\pi_k(x,z)=\1\{\tau_z(x)\ge k\}$ and define the
group-specific manipulation buffer
\begin{equation}
  \rho_z=\sqrt{\frac{2}{\lambda_z}}\|\beta\|_2.
  \label{eq:rho_definition}
\end{equation}

\subsection{Closed-form Strategic Treatment}
\label{subsec:closed_form_strategic_treatment}

\begin{theorem}[Closed-form strategic treatment]
\label{thm:closed_form_access}
For $\pi_k$, the group-$z$ access burden is
\begin{equation}
  b_{k,z}(x)=\frac{\lambda_z}{2}
    \left(\frac{(k-\tau_z(x))_+}{\|\beta\|_2}\right)^2.
  \label{eq:closed_form_b_kz}
\end{equation}
Hence the treatment rule after strategic response is
\begin{equation}
  m_{k,z}(x)=\1\{\tau_z(x)\ge k-\rho_z\}.
  \label{eq:linear_strategic_treatment}
\end{equation}
\end{theorem}

The common announced cutoff therefore becomes the group-specific
effective baseline cutoff $k-\rho_z$. The indicator
$m_{k,z}(x)$, the treatment rate, and the harmful-treatment
mass are all weakly decreasing in $k$, directly from
\eqref{eq:linear_strategic_treatment}.

For this policy class, write
\[
  G^S(k)=\E[\tau_Z(X)m_{k,Z}(X)],\qquad
  H_z(k)=\E[(-\tau_z(X))_+m_{k,z}(X)\mid Z=z],
\]
\[
  F_z(k)=B_z^\star-\E[\tau_z(X)m_{k,z}(X)\mid Z=z].
\]
A welfare threshold is any
$k_{\maxth}\in\argmax_{k\in\R}G^S(k)$, and a fair threshold
is any $k_{\fair}\in\argmin_{k\in\R}\max_zF_z(k)$,
whenever the respective optimizer exists. The pointwise-safe
threshold is defined by
\begin{equation}
  k_{\safe}=\inf\{k:\ m_{k,z}(x)=0
     \ \text{for every }z\text{ and every }x
     \text{ with }\tau_z(x)<0\}.
  \label{eq:safe_pointwise_definition}
\end{equation}
This pointwise requirement is stronger than $H_z(k)=0$ for
all groups, which may hold at a smaller threshold if the
population distribution has gaps in its CATE support.

\subsection{Homogeneous Costs}
\label{subsec:homogeneous_threshold_collapse}

\begin{theorem}[Homogeneous-cost threshold collapse]
\label{thm:homogeneous_collapse}
Suppose all groups have the same manipulation cost
parameter $\lambda>0$, and let
\[
  \rho=\sqrt{\frac{2}{\lambda}}\|\beta\|_2.
\]
Then the common threshold $k=\rho$ implements the
ideal CATE allocation in every group:
\[
  m_{\rho,z}(x)=\1\{\tau_z(x)\ge0\}.
\]
Consequently,
\[
  G^S(\rho)=G^\star,\qquad
  H_z(\rho)=F_z(\rho)=0
  \quad\text{for all }z\in\Z.
\]
Thus $k=\rho$ is simultaneously welfare-maximizing,
pointwise-safe, and minimax group-regret optimal.
If the welfare and minimax group-regret minimizers
are unique, then
\[
  k_{\maxth}=k_{\safe}=k_{\fair}=\rho.
\]
\end{theorem}

\subsection{Heterogeneous Costs and Policy Tradeoffs}
\label{subsec:heterogeneous_welfare_safety}

Write $\rho_{\max}=\max_z\rho_z$ and
$\rho_{\min}=\min_z\rho_z$.

\begin{theorem}[Pointwise-safe common threshold]
\label{thm:no_exploit_threshold}
Under the linear--quadratic model above, the smallest common
threshold excluding every negative-CATE baseline type from
treatment is
\[
  k_{\safe}=\rho_{\max}.
\]
At this threshold, $H_z(k_{\safe})=0$, and the regret in
group $z$ is
\begin{equation}
  F_z(k_{\safe})=
  \E\!\left[\tau_z(X)
  \1\{0<\tau_z(X)<\rho_{\max}-\rho_z\}\mid Z=z\right].
  \label{eq:safe_group_regret}
\end{equation}
\end{theorem}

The common safe threshold is governed by the most manipulable
group. Groups with smaller buffers can lose positive-CATE
applicants who cannot reach this common threshold. If
group-specific announced thresholds are permitted,
$k_z=\rho_z$ instead gives
\begin{equation}
  m_{k_z,z}(x)=\1\{\tau_z(x)\ge0\},\qquad
  G^S=G^\star,\quad F_z=H_z=0.
  \label{eq:group_specific_correction}
\end{equation}
This follows directly from the closed-form treatment rule.

The welfare threshold depends on treatment-effect distributions
as well as manipulation buffers. Let $p_z=\Pbb(Z=z)$, and let
$f_z$ denote the density of $\tau_z(X)$ given $Z=z$.

\begin{theorem}[Necessary condition for welfare optimality]
\label{thm:kmax_foc}
Suppose each $f_z$ is continuous at the effective cutoffs
$k_{\maxth}-\rho_z$, and a finite welfare maximizer
$k_{\maxth}$ exists. Then
\begin{equation}
  \sum_{z\in\Z}p_z(k_{\maxth}-\rho_z)
  f_z(k_{\maxth}-\rho_z)=0.
  \label{eq:kmax_foc}
\end{equation}
The condition is necessary, but not generally sufficient.
\end{theorem}

If the conditional densities take a common positive value
at those effective cutoffs, the condition reduces to
$k_{\maxth}=\sum_zp_z\rho_z$.

The fair threshold depends on the group distributions too.
Using~\eqref{eq:linear_strategic_treatment}, group regret is
\begin{equation}
\begin{aligned}
  F_z(k)=\E\bigl[& (\tau_z(X))_+
    \1\{\tau_z(X)<k-\rho_z\}\\
   &+(-\tau_z(X))_+
    \1\{\tau_z(X)\ge k-\rho_z\}
    \mid Z=z\bigr].
\end{aligned}
  \label{eq:group_regret_threshold}
\end{equation}

\begin{theorem}[Unavoidable effective-cutoff disparity]
\label{thm:cutoff_disparity}
For any common announced threshold,
\begin{equation}
  \inf_{k\in\R}\max_{z\in\Z}|k-\rho_z|
  =\frac{\rho_{\max}-\rho_{\min}}{2},
  \label{eq:cutoff_disparity}
\end{equation}
with a minimizer
$k=(\rho_{\max}+\rho_{\min})/2$.
Thus heterogeneous buffers preclude making every effective
cutoff equal to zero with one common threshold.
\end{theorem}

This bound concerns nominal cutoff disparity, not welfare
regret: a cutoff mismatch need not produce positive regret
when the population has no treatment-effect mass in the
relevant interval. For example, with $\rho_L=0.90$ and
$\rho_H=0.35$, the common safe threshold $k=0.90$
imposes effective cutoff $0.55$ on the higher-cost group.
The cutoff midpoint $k=0.625$ minimizes the largest absolute
cutoff error, but need not minimize worst-group welfare regret.

\section{Estimation and Policy Learning}
\label{sec:estimation}

The results of Section~\ref{sec:linear_l2} characterize policy
thresholds in terms of treatment effects and manipulation buffers.
We now consider how those thresholds are selected when the CATE
must be learned from data and the costs of strategic response may
be unknown. The distinction matters because a correction that is
exact for a known CATE score need not be safe when the announced
score is estimated.

\subsection{Estimating Treatment Effects and Strategic Responses}

Suppose a first-stage CATE estimator has the affine form
$\widehat\tau_z(x)=\widehat\beta^\top x+\widehat\delta_z$.
In our observational design, this estimator is fitted using
baseline covariates, group membership, treatment, and outcomes,
under the identification conditions stated in Section~\ref{sec:model}.
For known quadratic cost parameters $\lambda_z$, the exact
manipulation buffer for the \emph{estimated announced score} is
\begin{equation}
  \widehat\rho_z
  =\sqrt{\frac{2}{\lambda_z}}\|\widehat\beta\|_2.
  \label{eq:estimated_buffer}
\end{equation}
The dependence on $\widehat\beta$ is important: even with known
costs, the response buffer corresponds to the score that agents
actually face, not necessarily to the true CATE score.

When the cost parameters are unknown, behavioral responses to a
known, fixed eligibility rule can instead inform the buffer.
In the deterministic model, an initially ineligible applicant
obtains treatment exactly when the distance to the announced
score threshold is within the group's response buffer. Observed
successful and unsuccessful responses therefore give lower and
upper bounds, respectively. Such bounds identify the
buffer for the rule under which the responses were collected;
they cannot automatically be transferred to a different announced
score. The deterministic approximation can also fail when
responses are noisy or private treatment values vary across
applicants, as examined in Section~\ref{subsec:learning_experiments}.

\subsection{Selecting Thresholds with Estimated Quantities}

Let $\widehat\rho_z$ denote either the buffer computed from known
costs in~\eqref{eq:estimated_buffer} or an estimate obtained from
response data for the same announced score. The estimated
strategic treatment indicator is
\begin{equation}
  \widehat m_{k,z}(x)
  =\1\{\widehat\tau_z(x)+\widehat\rho_z\ge k\}.
  \label{eq:plugin_treatment}
\end{equation}
For threshold selection, consider a sample
$\{(X_i,Z_i)\}_{i=1}^n$ with at least one observation from
each group. Write $\widehat\tau_i=\widehat\tau_{Z_i}(X_i)$ and
set
\begin{equation}
  a_i=\widehat\tau_i+\widehat\rho_{Z_i}.
  \label{eq:estimated_breakpoints}
\end{equation}
Then $\widehat m_i(k)=\1\{a_i\ge k\}$, and the empirical
strategic welfare and group welfare regret are
\begin{equation}
\begin{aligned}
  \widehat G^S(k)
    &=\frac{1}{n}\sum_{i=1}^n
      \widehat\tau_i\1\{a_i\ge k\},\\
  \widehat F_z(k)
    &=\frac{1}{n_z}\sum_{i:Z_i=z}
      \bigl[(\widehat\tau_i)_+
        -\widehat\tau_i\1\{a_i\ge k\}\bigr],
\end{aligned}
\label{eq:empirical_policy_objectives}
\end{equation}
where $n_z$ is the number of selection observations in group $z$.
These objectives use the estimated CATE to evaluate which baseline
types benefit from treatment, and the estimated buffer to determine
which types can reach the announced treatment region.

Both empirical objectives are constant between consecutive values
of $a_i$. Thus the welfare and minimax-regret thresholds can be
found by checking the distinct breakpoints, together with any
threshold above $\max_i a_i$ to include the no-treatment allocation.
Explicitly, over these candidate thresholds, define
\[
  \widehat k_{\maxth}\in\argmax_k\widehat G^S(k),
  \qquad
  \widehat k_{\fair}\in\argmin_k\max_{z\in\Z}\widehat F_z(k).
\]
Sorting the breakpoints permits the required sums to be updated
as applicants enter the treatment region. Sample splitting can
separate first-stage CATE fitting from threshold selection,
but is not required for this finite candidate search.

When the response buffers of the estimated score are known,
the smallest common threshold that is pointwise safe \emph{relative
to the estimated CATE} is
\[
  \widehat k_{\safe}=\max_{z\in\Z}\widehat\rho_z.
\]
If buffers are learned from deterministic response data, using
valid upper bounds instead of point estimates guards against
underestimating strategic reachability. Neither correction by
itself controls errors in the estimated treatment effects.
Appendix gives a conditional theorem
for pointwise safety relative to the true CATE when valid uniform
CATE-error and buffer bounds are available.

\section{Experiments}
\label{sec:experiments}

\subsection{Policy Tradeoffs under Strategic Response}
\label{subsec:policy_tradeoff_experiments}

We first examine the structural tradeoffs from
Section~\ref{sec:linear_l2} when manipulation costs are known but
treatment effects must be estimated. There are three groups with
probabilities $0.60$, $0.25$, and $0.15$. Baseline covariates
follow $X\sim N(0,I_5)$, and the true group-conditional CATE is
$\tau_z(x)=\beta^\top x+\delta_z$. Treatment is assigned
observationally with propensities depending on $X$ and $Z$;
Appendix specifies the outcome
model, propensity score, and CATE parameters.

The group-specific manipulation buffers are
$\rho_{\text{low}}=0.90$, $\rho_{\text{medium}}=0.70$, and
$\rho_{\text{high}}=0.35$. Thus individuals in the low-cost group
can cross a given score threshold more easily than those in the
high-cost group. We estimate the affine CATE using observational
data and compute the exact response buffers for its announced
score from the known cost parameters. We then compare the
uncorrected threshold $k=0$, the plug-in welfare threshold
$\widehat k_{\maxth}$, the common safe correction
$\widehat k_{\safe}$, and the minimax group-regret threshold
$\widehat k_{\fair}$. The group-specific correction
$k_z=\widehat\rho_z$ is also included as a benchmark for
policies that need not share one numerical threshold.

\begin{figure}[t]
\centering
\includegraphics[width=0.94\linewidth]{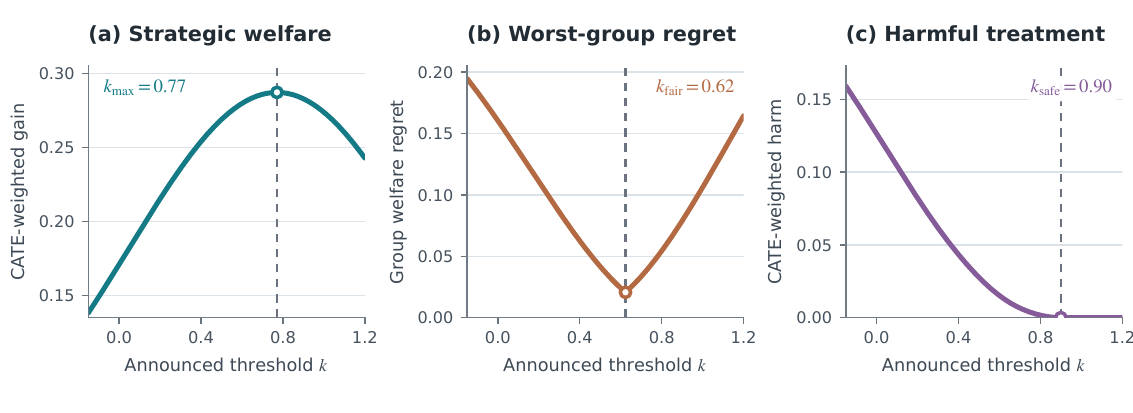}
\caption{Population threshold tradeoffs under the three-group
Gaussian design. (a) Strategic welfare $G^S(k)$, (b) worst-group
welfare regret $\max_z F_z(k)$, and (c) aggregate harmful-treatment
mass $H(k)$. Each vertical line marks the optimum for its
corresponding objective (the smallest pointwise-safe threshold
in panel (c)).}
\label{fig:thresholds_and_tradeoffs}
\end{figure}

Figure~\ref{fig:thresholds_and_tradeoffs} shows how the
\emph{population} objectives vary with the common announced
threshold. The panels separate strategic welfare, worst-group
welfare regret, and aggregate harmful treatment so that each
objective can be read on its own scale. Increasing $k$ reduces
harmful treatment but eventually excludes types with positive
CATEs, especially in groups with smaller buffers. The maximizing
welfare, minimizing worst-group regret, and smallest pointwise-safe
thresholds differ in this design. These are population curves
implied by the Gaussian data-generating process, not additional
fitted-policy Monte Carlo estimates.

Table~\ref{tab:plugin_aggregate} evaluates \emph{estimated}
policies using the true CATE after strategic response. It reports
Monte Carlo means over 200 replications, each with 6,000 causal
training observations and an independent test sample of 50,000.
Regret relative to the ideal positive-CATE allocation is divided
into harmful treatment and missed beneficial treatment, alongside
aggregate welfare, worst-group regret, and treatment rates.

\begin{table}[t]
\centering
\caption{Plug-in policy performance over 200 Monte Carlo
replications with estimated CATEs and known manipulation costs.}
\label{tab:plugin_aggregate}
\small
\setlength{\tabcolsep}{2.8pt}
\begin{tabular}{@{}lccccccc@{}}
\toprule
Policy & $\widehat{k}$ & $G^S$ & Regret & Harmful
 & Missed & $\max_z F_z$ & Treat rate \\
\midrule
Naive $k=0$ & 0.000 & 0.163 & 0.135 & 0.135
 & \textbf{0.000} & 0.168 & 0.867 \\
Welfare $k_{\maxth}$ & 0.773 & \textbf{0.285}
 & \textbf{0.012} & 0.005 & 0.007 & 0.044 & 0.523 \\
Safe $k_{\safe}$ & 0.901 & 0.283 & 0.014
 & \textbf{0.000} & 0.014 & 0.073 & 0.454 \\
Fair $k_{\fair}$ & 0.625 & 0.277 & 0.020
 & 0.017 & 0.002 & \textbf{0.027} & 0.602 \\
Group-specific $k_z$ & -- & \textbf{0.297}
 & \textbf{0.001} & \textbf{0.001}
 & \textbf{0.000} & \textbf{0.003} & 0.526 \\
\bottomrule
\end{tabular}
\end{table}

The naive threshold treats many applicants who can strategically
enter the treatment region despite having negative treatment
effects. The welfare-maximizing common threshold reduces this
harm while retaining most beneficial treatment, achieving the
highest aggregate welfare among the common-threshold policies.
The safe threshold reduces harmful treatment further but misses
more beneficial applicants, especially in groups whose smaller
buffers make the common cutoff harder to reach. The fairness
threshold reduces the largest group welfare loss at a modest cost
to aggregate welfare. The group-specific correction nearly
recovers the ideal allocation by offsetting each group's buffer
separately. These comparisons agree with the structural
predictions, but group-specific eligibility rules may face
institutional or legal restrictions.

\subsection{Estimating Responses and Learning Policies}
\label{subsec:learning_experiments}

We next examine the additional uncertainty that arises when the
manipulation buffers are estimated from behavioral responses.
The response-estimation study is separate from the policy
comparison above. Across 250 replications per setting, a
deterministic response model yields a valid score-distance
bracket with midpoint RMSE $0.0033$. In contrast, under logistic
response noise or heterogeneous private treatment values,
successes and failures can overlap in score distance and no
single deterministic cutoff need fit the data.

We also evaluate downstream policy performance when both the
CATE and the response buffer must be estimated. This separate
end-to-end design uses samples of size 6,000 for causal
estimation, 1,200 for response estimation, 10,000 for threshold
selection, and 50,000 for evaluation, over 200 replications.
The welfare-oriented policy uses the midpoint of each estimated
response-buffer bracket. The safety-oriented policy uses its
upper endpoint instead, guarding against underestimating how
far applicants can move relative to the estimated score.
Both policies use an estimated CATE and are evaluated against
the true treatment effects.

\begin{table}[t]
\centering
\caption{End-to-end policy performance over 200 Monte Carlo
replications with estimated CATEs and manipulation buffers.
The safety-oriented policy uses upper buffer bounds.}
\label{tab:end_to_end}
\small
\begin{tabular}{lccc}
\toprule
Policy & Regret & Harm & Missed benefit \\
\midrule
Uncorrected naive rule & 0.1278 & 0.1278 & 0.0000 \\
Welfare, midpoint buffer & 0.0118 & 0.0036 & 0.0082 \\
Safety, upper buffer & 0.0159 & 0.00036 & 0.0155 \\
\bottomrule
\end{tabular}
\end{table}

Table~\ref{tab:end_to_end} shows that accounting for strategic
response substantially reduces regret even with estimated
buffers. Using upper response-buffer bounds reduces harmful
treatment further, though it increases missed beneficial
treatment. Importantly, harmful treatment is not exactly zero:
conservative estimation of reachability alone cannot prevent
errors in the sign of the true CATE.

\section{Discussion}
\label{sec:discussion}

This paper studies strategic causal policy learning, where agents respond strategically to an announced policy before treatment is assigned, but causal welfare is evaluated using baseline treatment effects. 
The resulting mismatch is simple: assignment depends on manipulated covariates \(X^\pi\), while treatment gain remains the baseline CATE
(in the group-conditional analysis, \(\tau_Z(X)\)). 
Thus strategic response is not only a distribution-shift or classification issue; it changes causal welfare by changing which baseline types can access treatment.

We represent this mismatch through an access-burden indicator \(m_\pi\), derive a manipulation-gap identity, and decompose regret into harmful treatment and missed beneficial treatment. 
Welfare maximization, harmful-treatment control, and minimax group welfare regret are possible objectives in this setting; CATE thresholds provide one tractable policy class. 
In linear CATE models with squared \(L_2\) manipulation costs, homogeneous costs admit a single buffer correction, while heterogeneous costs make these thresholds diverge. 
The experiments illustrate those threshold tradeoffs and the sensitivity of learned policies to response and CATE estimation. This framework may help institutions design treatment rules that better account for harmful treatment and unequal access burden under strategic eligibility. At the same time, incorrect cost models or inappropriate use of group-specific thresholds could worsen exclusion or unfairness, so deployment requires application-specific legal, ethical, and empirical validation.

Several limitations remain. 
We focus on pure gaming, where manipulation affects eligibility but not potential outcomes; in some settings, strategic actions may also be productive investments or interventions that change outcome-relevant states \citep{karimi2021algorithmic}. While the base structural results use a known cost model, Section~\ref{sec:estimation} studies what can be inferred from response data under additional assumptions. Arbitrary nonlinear costs, constrained actions, and valid uniform CATE confidence bands remain demanding identification and inference problems. Finally, our strongest closed-form results rely on linear CATEs and squared \(L_2\) costs. Extending the framework to nonlinear CATE models, learned cost functions, and institutional constraints on group-specific thresholds is an important direction for future work.

\bibliographystyle{plainnat}
\bibliography{references}

\end{document}